\pdfoutput=1 
\documentclass{aa}
\usepackage{ulem}
\usepackage{array}
\usepackage{graphicx}
\usepackage{txfonts}
\usepackage{natbib}	
\usepackage{hyperref}
\graphicspath{{./plots/}}
\begin{document} 


   \title{Interferometric Evidence for Quantum Heated Particles in the Inner Region of Protoplanetary Disks around Herbig Stars}

   \author{L. Klarmann \inst{1}
          \and
          M. Benisty \inst{2}
          \and
          M. Min \inst{3,1}
          \and
          C. Dominik \inst{1}
          \and
          J.-P. Berger \inst{4}
          \and
          L. B. F. M. Waters \inst{3,1}
          \and
          J. Kluska \inst{5}
          \and
          B. Lazareff \inst{2}
          \and
          J.-B. Le Bouquin \inst{2}
          }

   \institute{Astronomical Institute Anton Pannekoek, University of Amsterdam, Science Park 904, 1098 XH Amsterdam, The Netherlands\\
              \email{l.a.klarmann@uva.nl}
         \and
             Universit\`e Grenoble Alpes, IPAG, 38000 Grenoble, France; CNRS, IPAG, 38000 Grenoble, France
         \and
             SRON Netherlands Institute for Space Research, Sorbonnelaan 2, 3584 CA Utrecht, The Netherlands
         \and
         	 European Southern Observatory, Karl Schwarschild Strasse, 2, Garching bei M\"unchen, Germany 
         \and
         	 University of Exeter, School of Physics, Stocker Road, Exeter, EX4 4QL, UK 
             }

    \date{Received April 29, 2016; accepted December 16, 2016}

  \abstract
   {To understand the chemical composition of planets, it is important to know the chemical composition of the region where they form in protoplanetary disks. Due to its fundamental role in chemical and biological processes, carbon is a key element to trace.}
  {We aim to identify the carriers and processes behind the extended NIR flux observed around several Herbig stars.}
   {We compare the extended NIR flux from objects in the PIONIER Herbig Ae/Be survey with their flux in the PAH features. HD$\,$100453 is used as a benchmark case to investigate the influence of quantum heated particles, like PAHs or very small carbonaceous grains, in more detail. We use the Monte Carlo radiative transfer code MCMax to do a parameter study of the QHP size and scale-height and examine the influence of quantum heating on the amount of extended flux in the NIR visibilities.}
   {There is a correlation between the PAH feature flux of a disk and the amount of its extended NIR flux. We find that very small carbonaceous grains create the observed extended NIR flux around HD$\,$100453 and still lead to a realistic SED. These results can not be achieved without using quantum heating effects, e.g. only with scattered light and grains in thermal equilibrium.}
   {It is possible to explain the extended NIR emission around Herbig stars with the presence of carbonaceous, quantum heated particles. Interferometric observations can be used to constrain the spatial distribution and typical size of carbonaceous material in the terrestrial planet forming region.}

\titlerunning{Interferometric Evidence for QHPs in the Inner Region of PPDs around Herbig Stars}
   \maketitle
%
\section{Introduction}

Protoplanetary disks are the birth places of planets.  One of the most important goals of current research into exoplanets and
their formation is to trace the properties of protoplanetary disks and understand their impact on the resulting planetary system.  This applies both to the architecture of the planetary systems - what kind of planets will be formed, where will they be located once the disk has disappeared - as well as the chemical properties of planets.  What will be the composition of planets and their atmospheres, and how does this composition relate to the chemistry in the disk or even to the composition of the primordial cloud from which the the planetary system formed.  Ultimately, we want to understand both the bulk composition of planets, the composition of planetary atmospheres, and the delivery of prebiotic molecules.
\\
To move this research along, it is important to trace the reservoirs of key elements in disks, and to trace these reservoirs not only in the outer disk, but also in the inner disk where terrestrial planets \citep[e.g.][]{Morbidelli12} or systems of super earth planets like the compact Kepler systems \citep{Lissauer11, Lissauer13} are born. A key element to be traced is carbon, not only because it has a large biological relevance \citep{Pace01}, but also due to the importance of the carbon to oxygen ratio. This ratio is a determining factor for the chemical evolution of disks \citep[e.g.][]{Oeberg11}, the bulk composition of planets \citep[e.g.][]{Moriarty14} and planetary atmospheres \citep[e.g.][]{Madhusudhan12}.
\\
What fractions of carbon are in solids that will readily take part in aggregation processes and thus become building blocks of terrestrial planets and the cores of Neptune-like planets or giants?  What carbon-bearing compounds remain in the gas phase so that they are less prone to end up in solid planets or cores?
\\
This is particularly interesting in the context of heavy depletion of carbon in the Earth crust \citep{Allegre01}, where the silicon-to-carbon ratio is four orders of magnitude lower than in the Sun, and still one order of magnitude lower than in meteorites \citep{Lee10, Pontoppidan14}.  Apparently, much of the carbon either stays in the outer disk in the form of ices \citep[e.g.][]{Qi04, Jorgensen05}, or is pushed into the gas phase where it will not be incorporated into solid planets or cores \citep[][and references therein.]{Lee10}. Atomic carbon and carbon-bearing molecules like CO and CO$_{2}$ can be traced through molecular line emission, \citep[e.g.][]{Bruderer12, Kama16} and with the advent of ALMA, there are finally tools available to do so also with sufficient spatial resolution to reach into the inner regions of protoplanetary disks \citep{ALMA15}.
Very small carbonaceous particles are hard to trace in disks because, unlike silicates and some oxides \citep{Koike06,Suto06}, they barely have specific spectroscopic signatures \citep{Draine84}. In the diffuse interstellar medium, observations of the 3.4$\,\mu$m absorption feature indicates the presence of hydrogenated amorphous carbon \citep{Pendelton02,Rawlings03}. These observations are in agreement with laboratory experiments on interstellar carbon dust analogues \citep[e.g.][]{Dartois04} and dust models \citep[e.g.][]{Jones13}. Some of this hydrogenated amorphous carbon will survive in the protoplanetary disk \citep{Pendelton02}, but the properties of carbonaceous particles in protoplanetary disks will in general differ from the properties in the ISM \citep{Apai10}.
\\
Large molecules (up to several hundred carbon atoms) like polycyclic aromatic hydrocarbons (PAHs) can be identified in protoplanetary disks in two ways: (i) by their CH and CC bond bending and stretching features \citep{Leger84, Allamandola85} and (ii) by the fact that they can be quantum heated \citep{Purcell76,Draine01} and are not in thermal equilibrium, a property that can be used to identify such grains by warm emission that is spatially much more extended than grains in thermal equilibrium would allow.
\\
PAHs have been observed in many protoplanetary disks around Herbig stars using method (i) \citep{Meeus01, Acke04, Keller08, Acke10}, but their spatial distribution is not well constrained: When both the continuum and PAH emission can be resolved by long slit spectroscopy, \citet{vanBoekel04} and \citet{Visser07} find that the PAH flux is more extended then the continuum flux, due to the effects of quantum heating. \citet{Habart06} confirm this result also for the 3.3 $\mu$m feature. \citet{Geers07} find that some of their sources are not extended in the PAH features, indicating that in these sources PAHs are confined to the innermost regions. \citet{Lagage06} and \citet{Doucet07} show that for HD\,97048 the emission in the PAH features at 8.6$\,\mu$m and 11.3$\,\mu$m follows the large-scale disk geometry. \citet{Maaskant14} determine the ionisation degree of the PAH emission from disks that have a gap, and conclude that ionised PAHs are located in the gap regions.
\\
With increasing particle size, the strength of the emission features decreases relative to the continuum emission, so it becomes hard to identify these particles using emission features. \citet{Berne09} attempt to disentangle the emission of PAHs and larger quantum heated particles (called VSGs in \citet{Berne09}) by the application of feature templates and show that a component of such larger quantum heated particles (QHPs) is necessary to understand the spectra of protoplanetary disks. However, the use of additive templates does not yield any spatial information. 
\\
In this paper we intend to show that a number of protoplanetary disks do show a component of extended emission that can only be explained by the presence of such quantum heated particles, and that those particles are larger than PAHs. Interferometric observations of the innermost regions of protoplanetary disks therefore offer a way to identify and trace a solid component of carbon in such disks, in the terrestrial planet-forming region. We make not effort to fit feature shapes in detail, but use them only as a limit on the contribution of our QHPs to such features. Instead, we focus on the spatial information from interferometric observations to identify QHPs by the spatial extension for the continuum emission (method (ii)).
\\
In Sec.~\ref{sec:PIONIER Data}, we present the HD$\,$100453 observations. In Sec.~\ref{sec:QHPs} we introduce quantum heated particles (QHPs) and show that their presence correlates with the observed extended flux around Herbig stars. We explain the radiative transfer models in Sec.~\ref{sec:RTM}. The results of the parameter study are presented and discussed in Sec.~\ref{sec:ResultsDiscussion}. We present our conclusions and further perspectives in Sec.~\ref{sec:Conclusions}. In App.~\ref{app:denstemp} we show examples for the density and temperature structure of our models.
%
\section{PIONIER Data}
\label{sec:PIONIER Data}
\begin{figure}
\centering
    \includegraphics[width=1.0\columnwidth]{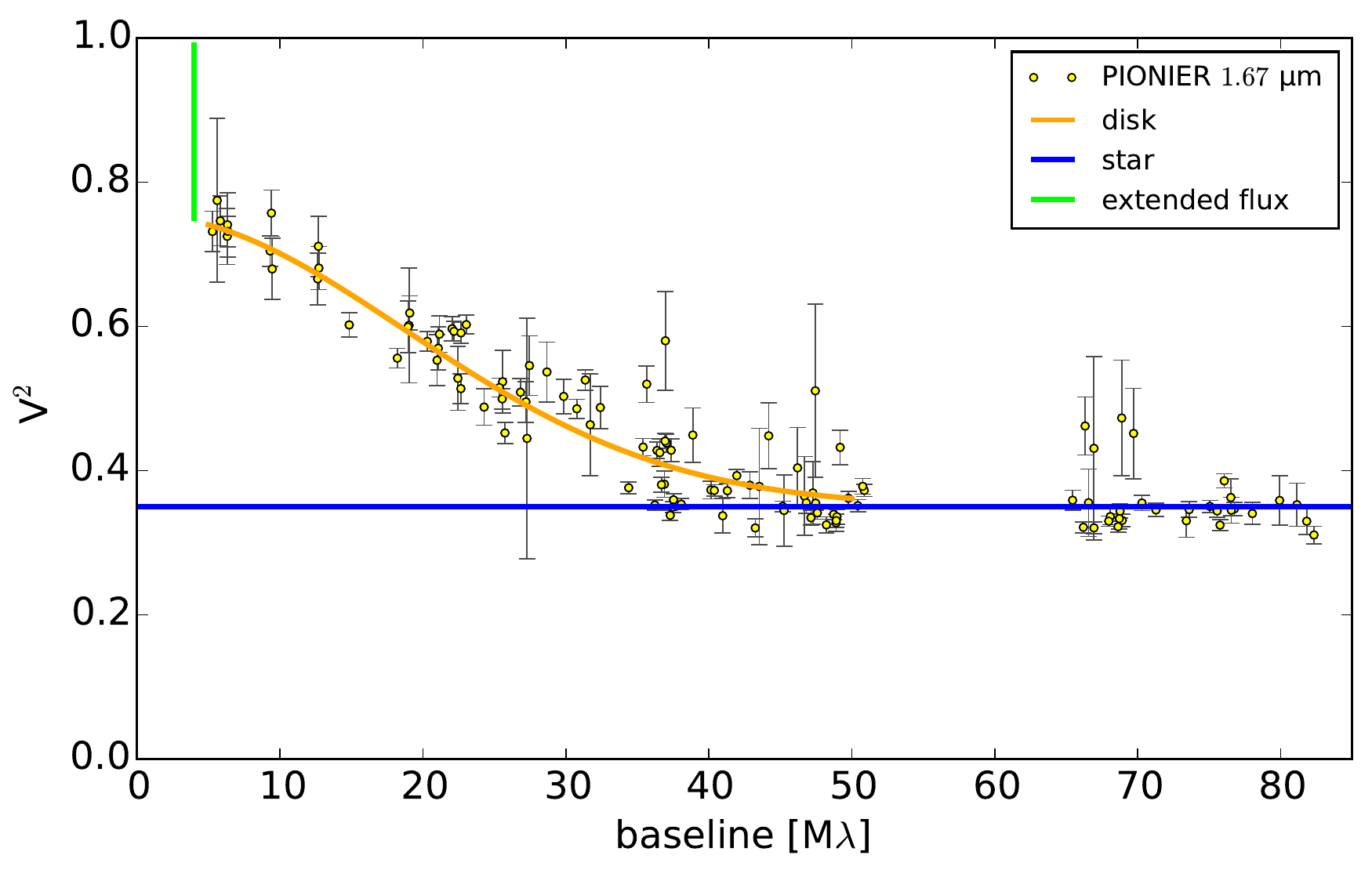}
     \caption{Squared, normalised visibilities $V^2$ against baseline in M$\mathrm{\lambda}$. The yellow circles show the PIONIER observations of HD$\,$100453 at 1.67$\,\mathrm{\mu}$m. We demonstrate a basic modelling approach that explains the data using a three different regimes. 
      }
      \label{fig:data_cartoon}
\end{figure}
\subsection{Observations}
\label{sub:Observations}
The interferometric data were obtained with the PIONIER instrument \citep{LeBouquin11, LeBouquin12} at the VLTI \citep{Merand14}. The observations that we consider in this paper were gathered as part of a PIONIER Large Program (190.C-0963, PI: Berger) from December 2012 and June 2013 \citep{Lazareff16, Kluska16} using three different 4-ATs configurations. The magnitude limit of the survey is at $m_H\leqslant 8$, but it does not meet any completness criteria. The Herbig Ae/Be stars within the survey are selected based on \citet{The93}, \cite{Malfait98} and \citet{Vieira03}. A list of all objects can be found in \citet{Lazareff16}. Each disk observation was preceded and followed by observations of a calibrator star\footnote{using SearchCal} in order to calibrate the instrumental transfer function. The data were reduced and calibrated with the pndrs package \citep{LeBouquin11}. The typical accuracy on the squared visibilities ($V^2$) is 5\%.\\
Fig.~\ref{fig:data_cartoon} shows the $V^2$ data of HD\,100453 at 1.67$\,\mathrm{\mu}$m against the baseline in M$\mathrm{\lambda}$ (yellow circles). 
\subsection{Interpretation}
\label{sub:Interpretation}

We start with a basic description of the visibility curve of HD$\,$100453. As shown in Fig.~\ref{fig:data_cartoon}, there are three different regimes. At very long baselines ($B>60\,$M$\lambda$), the $V^{2}$ are constant and provide the 1.67$\,\mathrm{\mu}$m stellar flux contribution to the visibilities (blue line). At intermediate baselines $8<B<50\,$M$\lambda$) we detect the bulk of the disk emission at 1.67$\,\mathrm{\mu}$m coming from regions at blackbody temperatures of ~1500 K (orange line), while at short baselines ($B<8\,$M$\lambda$), we probe extended emission that causes the $V^{2}$ to drop quickly from 1 to ~0.75 at these baselines (green line). Image reconstruction of HD$\,$100453 is still in progress (Kluska et al. in prep.) but preliminary results \citep{Kluska14} give an overview over the spatial flux distribution.\\ 
$V$ is normalised by the flux within the PIONIER field of view. Under typical atmospheric conditions (0.8$''$ seeing in the visible) the FWHM of the Gaussian profile at the focus of each telescope is of the order of 0.4$''$. For HD$\,$100453 at a distance of 114$\,$pc the flux emitted beyond 23$\,$au will not contribute to the interferometric measurement. The size of the maximum observable scale depends on the flux distribution and model parameters. A rough estimate would be a Gaussian distribution with a half intensity radius of 1$\,$au. With the baseline range going up to 82$\,$M$\lambda$, the resolution of the observation is 0.2$\,$au.\\
To show HD\,100453 in the context of the survey, we make use of parameters calculated by \citet{Lazareff16}, who present the PIONIER survey results and use geometric models to obtain structural parameters for each object. They define $f_{\text{h}}$ as the fraction of H band flux coming from the extended emission, $f_{\text{s}}$ as the fraction of H band flux coming from the star and $f_{\textbf{c}}$ as the flux attributed to the emission from the circumstellar disk. Hence $f_{\text{h}}+f_{\textbf{c}}+f_{\text{s}}=1$. The values for $f_{\text{h}}$ and $f_{\text{s}}$ can be found in Tab.~\ref{tab:fluxes}\\.

%
\section{Quantum Heated Particles}
\label{sec:QHPs}
\begin{figure}
\centering
    \includegraphics[width=1.0\columnwidth]{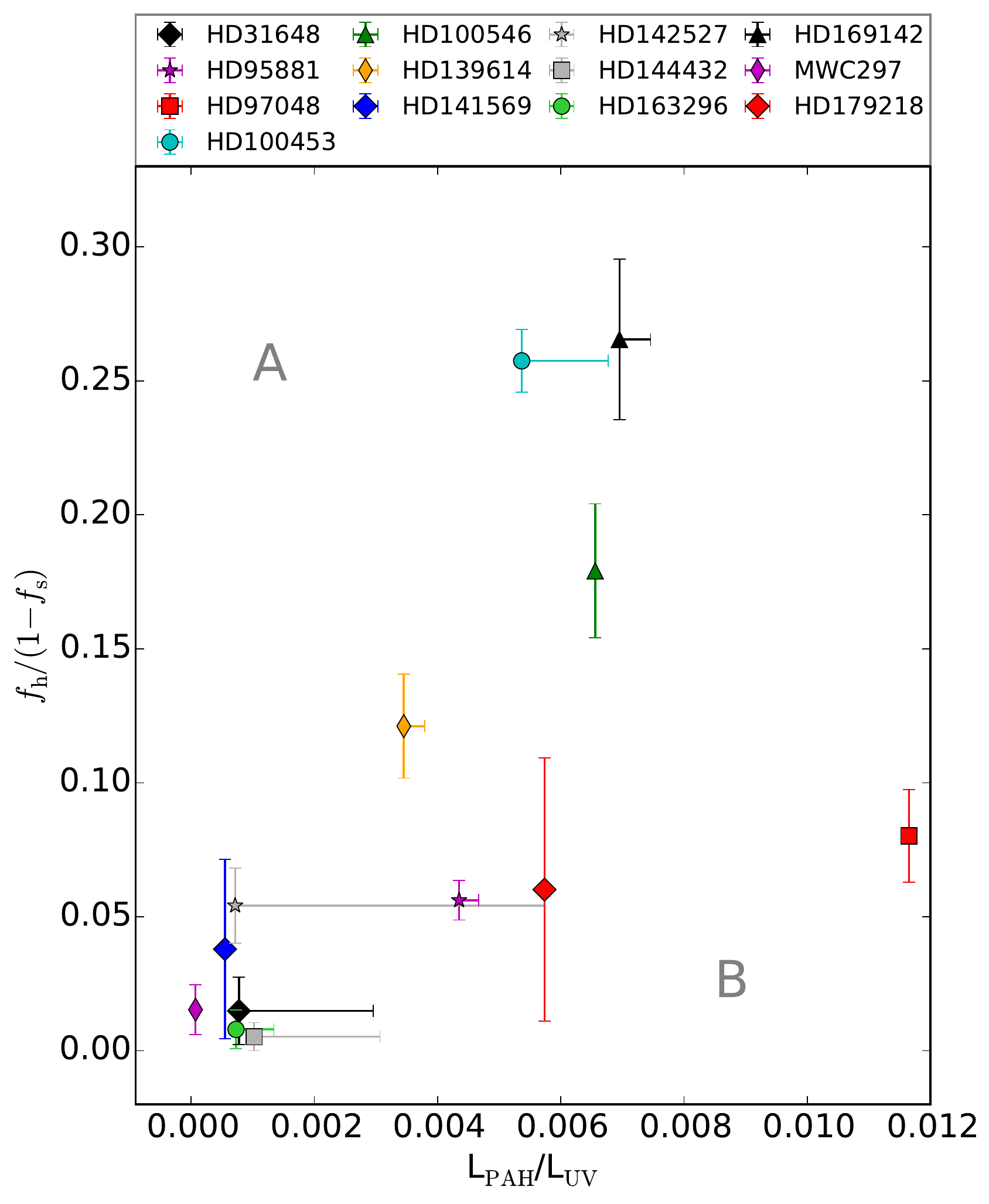}
     \caption{Extended flux as calculated from the PIONIER survey \citep{Lazareff16} plotted against the normalised PAH luminosity from the ISO observations \citep{Acke04}. For most disks, only an upper limit is given for one ore more of the PAH features. In that case, the symbol indicates the actually observed luminosity while the horizontal errorbar indicates the luminosity including upper limits. The legend shows the name of each source.
      }
      \label{fig:UV_drop_scatter}
\end{figure}

The large amount of extended flux from HD$\,$100453 means that a significant part of the 1.67$\,\mathrm{\mu}$m emission does not come from the inner rim, but is extended to at least a few au. That far from the star, the dust in the disk is not hot enough to emit thermally at this wavelength.\\ 
In T Tauri stars, the extended emission has been attributed to scattered light \citep[][in the following P08; \citealp{Anthonioz15}]{Pinte08}. Since T Tauri stars have a lower effective temperature than Herbig stars, they emit more strongly in the NIR. This light is then scattered on small grains in the surface layer of the protoplanetary disk. Due to the much higher temperature and thus much bluer color of the central star, it is not possible to get enough scattered light from full disks around Herbig stars. After significant modelling efforts, we do not find an explanation based on scattering. This is in agreement with P08.\\
Another source of extended NIR flux that can cause this $V^2$ drop at short baselines are quantum heated particles (QHPs) like PAHs or very small grains. Unlike conventional dust grains, QHPs are not in thermal equilibrium. Instead, a QHP absorbs a UV photon and is heated to very high temperatures.  It then cools down again very quickly by emitting photons in the NIR and stays cold until it is hit by the next photon. This mechanism, including multi-photon heating events and different ionisation states, is described in detail by \citet{Draine01}. A well known example for QHPs are PAHs, but the mechanism also works for very small grains.\\ 
To investigate if this is a promising approach, we examine if the extended NIR flux does correlate with the amount of QHPs in a disk. We collect the extended and stellar flux for each disk within the PIONIER survey from \citet{Lazareff16}. PAHs and very small grains both contribute to the PAH emission features in the MIR (see Fig.~\ref{fig:Spectrum_size} for the size dependence of the contribution), so the flux within the PAH features can be used as an indicator for the presence of quantum heated particles. We pick the disks from the PIONIER survey that overlap with the ISO survey \citep{Acke04}. This survey contains the flux of the PAH features at 3.3, 6.2, 7.7 and 8.6 $\,\mathrm{\mu}$m, but not for the 11.2 $\,\mathrm{\mu}$m feature, since it is blended with the 11.3 $\,\mathrm{\mu}$m silicate feature. As \citet{Acke04} for their Tab.~9, we use the sum of the fluxes in the four PAH features to calculate the PAH luminosity L$_{\text{PAH}}$ for each source. PAHs are excited mainly by UV photons. To remove the PAH flux variation introduced by the different stellar UV fluxes we normalize the PAH luminosity with the stellar UV luminosity L$_{\text{UV}}$. Since only 4 of our disks have measurements for all four PAH features, we also include disks where only upper limits were obtained for one, two or three of the PAH features.\\
In Fig.~\ref{fig:UV_drop_scatter}, the relative extended flux $f_{\text{h}}/(1-f_{\text{s}})$ (the overresolved part of the circumstellar emission) is plotted against the normalised PAH luminosity L$_{\text{PAH}}/$L$_{\text{UV}}$. The normalised luminosities for sources with four flux measurements are indicated only by a symbol. For disks that have one, two or three upper limits, we first assume that PAH features with the upper limit contain no flux. The L$_{\text{PAH}}/$L$_{\text{UV}}$ values obtained like that are also indicated by symbols. For these disks, we  calculate L$_{\text{PAH}}/$L$_{\text{UV}}$ a second time, including the upper limits as flux values. The result is plotted as the horizontal errorbar.\\ 
Fig.~\ref{fig:UV_drop_scatter} shows that disks with more extended flux have a larger normalised  PAH luminosity (indicating the presence of more PAHs). This is expected if particles emitting the PAH features also contribute significantly to the extended flux.\\
The upper left area (labelled A) and the lower right area (labelled B) are (nearly) empty. That means that no disk has a large amount of extended emission, but no PAHs (area A), and (nearly) no disk has a large amount of PAHs, but only a small amount of extended emission (area B). The one exception is HD$\,$97048 (red square). While this disk has the largest measured PAH flux, it shows only a moderate drop. I has PAH emission from the inner region \citep{Habart06}, but \citet{Maaskant14, vanBoekel04} showed that most of the PAHs in this disk are at a distance of more than 70$\,$au from the star, which is outside the field of view of the PIONIER instrument (see Sec.~\ref{sub:Interpretation}). \\
This is one reason for the large scatter of the disks: Depending on their distance from the star, the spectral type of the star and the distance of their PAHs from their central star, the fraction of PAHs that contributes flux to the ISO survey, but not the PIONIER observations, will naturally vary. HD$\,$95881 (magenta star) for example has only a very small dust disk, but a gas disk and PAH emission extending out to 200$\,$au \citep{Verhoeff10}. We therefore expect that its extended NIR flux is mainly caused by PAHs. \citet{Olofsson13} speculate on a possible influence of PAHs on the short baselines of PIONIER $V^2$ in their study of the Chamaleon-I region, but do not investigate it due to large uncertainties in PAH size and distribution.\\
Another reason for the scatter is the inhomogenity of the survey that contains disks during various evolutionary stages, including transitional disks.

\section{Radiative Transfer Modelling}
\label{sec:RTM}
\begin{table}
\caption{Model parameters used in each model. The parameters for the DIANA dust and QHP particles are separated by a horizontal line. The negative power law index indicates a rise in the surface density for larger radii. More description and references can be found in the text.}
\label{tab:parameters}
\centering
\begin{tabular}{lr}
\hline\hline
Parameter & Value \\
\hline
stellar mass   & 1.66$\,$M$_{\odot}$ \\
stellar temperature   & 7400$\,$K \\
stellar luminosity & 8.04$\,$L$_{\odot}$ \\
spectral type & A9Ve \\ 
distance & 114$^{+11}_{-9}\,$pc \\
\hline
inner radius & 0.27$\,$au \\
outer radius & 200$\,$au \\
gap area & 1$\,$au - 17$\,$au \\
\hline
dust mass & 3.2$\times10^{-4}\,$M$_{\odot}$\\
min. dust radius & 0.05$\,\mu$m \\
max. dust radius & 3 mm \\
size dist. power index & 3.5 \\
porosity & 0.25\\
mean opacity & 0.48\\
\hline
QHP mass & 1$\times10^{-9}\,$M$_{\odot}$\\
inner radius QHP & 0.5$\,$au \\
outer radius QHP & 17$\,$au \\
power law index change & 1$\,$au \\
inner surface density power law index & -1 \\
outer surface density power law index & 0 \\
\hline

\end{tabular}
\end{table}
\label{sec:RTM}
\begin{table*}
\caption{For each model discussed in this paper, we list the QHP size in Number of carbon atoms and the QHP scale-height at 1$\,$au in au. We also indicate whether a gap is present and if the quantum heating routine is turned on during the calculations. The resulting $V^2$ values at 5.3 M$\mathrm{\lambda}$ and 12.8 M$\mathrm{\lambda}$ can be found in the next two lines. Model S18C100K is closest to the data. In the last line we compare the flux of the 6.2$\,\mu$m feature for the model and SPITZER observations \citep{Acke10}}.
\label{tab:models}
\centering
\begin{tabular}{>{\raggedright}p{2.4cm}>{\centering}p{1.2cm}
>{\centering}p{1.4cm}>{\centering}p{1.4cm}>{\centering}p{1.4cm}>{\centering}p{1.5cm}>{\centering}p{1.5cm}>{\centering}p{1.8cm}c}
\hline\hline
Parameter & S18C250 & S18C100K & S18C500K & S18C1.6M & S10C100K & S35C100K & S18C100Kng & S18C100Knqh \\
\hline
size [\#C$\,$atoms]   & $10^{2}$ & $10^{5}$ & $5\times10^{5}$ & $1.6\times10^{6}$ & $10^{5}$ & $10^{5}$ & $10^{5}$ & $10^{5}$\\
scale-height [au] & $0.18$   & $0.18$   & $0.18$          & $0.18$            & $0.10$   & $0.35$   & $0.18$   & $0.18$\\
gap              & yes      & yes      & yes             & yes               & yes      & yes      & no       & yes      \\
quantum heating  & yes      & yes      & yes             & yes               & yes      & yes      & yes      & no      \\
$V^{2}$ at 5.3 M$\mathrm{\lambda}$
				 & 0.85     & 0.84     &	0.88 			& 0.88	   			& 0.90	  &	0.80		 & 0.87		& 0.95    \\
$V^{2}$ at 12.8 M$\mathrm{\lambda}$
				 & 0.75     & 0.65     &	0.70 			& 0.70	     		& 0.74	  &	0.54		 & 0.69		& 0.84    \\
F$_{\mathrm{model}}/$F$_{\mathrm{spitzer}}$at 6.2$\,\mu$m
				& 2.44	& 1.30     & 0.89			& 0.88 & & & &\\
\hline
\end{tabular}
\end{table*}
Since HD\,100453 shows a significant amount of extended flux and is observed at a large number of baselines (good uv-plane coverage), we use it as a benchmark to investigate the extended flux in more detail using radiative transfer modelling. We calculate eight models for a small parameter study to constrain the size and scale-height of the QHPs. Then we use the size and scale-height values of the best of the six models to show the influence of a gap in the disk and of the quantum heating. The name, parameters and resulting drop depth of each model can be found in Tab.~\ref{tab:models}. We use HD$\,$100453 as an example to show the influence of the QHPs. Therefore, our best model is a reasonable description, but not a 'best fit model' in the usual sense.\\
We use the radiative transfer code MCMax \citep{Min09}. It solves 3D radiative transfer \citep[see also][]{Bjorkman01} to calculate the 2D dust density and temperature structure of our disk setup. After the disk structure and temperature calculation has converged, MCMax calculates an image and the $V^2$ for different wavelengths and baselines.\\
Each model has the same basic setup. First we fix the stellar parameters, following \citet[][and references therein; in the following K16]{Khalafinejad16}. HD$\,$100453 is a 10$\,$Myr old Herbig star with spectral type A9Ve at a distance of 114$^{+11}_{-9}\,$pc. It has an effective temperature of 7400$\,$K, a mass of 1.66$\,$M$_{\odot}$ and a luminosity of 8.04$\,$L$_{\odot}$. Like K16, we implement the star using a Kurucz model. \\
For the conventional dust we take the DIANA standard dust properties as described in \citet{Woitke16}. Based on the K16 spectrum and Q-band fits we take a dust mass of 3.2$\times10^{-4}\,$M$_{\odot}$, distribute it according to a power law with index 1, use a gas to dust ratio of 100 and an outer disk radius of 200$\,$au. We also introduce a gap and set the inner radius of the outer disk to 17$\,$au. Recently, this gap has also been found in SPHERE observations by \citet{Wagner15}. The inner disk starts at $R_{\text{in}}=0.27\,$au, determined by the evaporation temperature of the dust grains ($\sim1450\,$K), and goes out to 1$\,$au.\\
Additionally, we fill the disk gap with with $10^{-9}\,$M$_{\odot}$  QHPs (with a flat surface density distribution) and investigate their influence on the depth of the drop by studying four different QHP sizes: 0.0008, 0.006, 0.01 and 0.015$\,\mathrm{\mu}$m, which corresponds to $10^{2}$, $10^{5}$, $5\times10^{5}$, $1.6\times10^{6}$ carbon atoms. In Tab.~\ref{tab:models}, the models can be found as S18C250, S18C100K, S18C500K and S18C1.6M. 
We use opacities from \citet{Li01} with updates from \citet{Draine07} for the QHPs and treat them during the temperature calculation as explained in Sec.~\ref{sec:QHPs}. For all but the S18C250, the QHPs are too large to fit the PAH category. For particles of this size, \citet{Li01} use a mixtures of graphite and PAHs for their opacities. In the NIR, the difference between these mixed opacties and graphite and amorphous carbon opacities are small. Our models could not be used to distinguish between them. We keep the name QHPs to indicate this ambiguity. In the MIR the mixed opacities still show some carbon features for the $10^{5}$ carbon atoms model, but they are extremly weak in the opacities of the two larger QHPs.\\
To avoid sharp edges in our disk setup, we create a continuous transition between the inner disk and the QHPs: we add a small amount of QHPs to the area from 0.5$\,$au to 1$\,$au. Their surface density follows a power law with an index of -1 (see also App.~\ref{app:denstemp}). All model parameters can also be found in Tab.~\ref{tab:parameters}.\\
Two examples (with and without QHPs) of the density and temperature distributions after the radiative transfer can be found in Fig.~\ref{fig:denstemp} in App.~\ref{app:denstemp}. The DIANA dust is in thermal equilibrium. Its temperature does not change over time. Only the region close to the star is it hot enough for the disk to emit thermally in the NIR. But when a QHP gets excited by a UV photon, it will emit a NIR photon during the excitation independently of the surrounding temperature. The area dominated by QHP emission can also be seen in Fig.~\ref{fig:denstemp}.
In MCMax, the dust scale-height is calculated via the midplane dust temperature. Since the QHPs are not in thermal equilibrium, this approach is not possible. We therefore fix the QHP scale-height at 0.18$\,$au from 1 to 17$\,$au for the first four models. Then we pick the best size model and change the scale-height of the QHPs to 0.10 and 0.35$\,$au. In Tab.~\ref{tab:models}, these models are called S10C100K and S35C100K.\\
To show that this approach also works in continuous disks, we remove the gap from model S18C100K and re-calculate it using a continuous disk, S18C100Kng. Finally, we turn off the quantum heating and treat the QHPs like small, conventional dust grains to see if they could possibly provide enough extended NIR flux via scattering (model S18C100Knqh).\\
Since it is sufficient to demonstrate the influence of QHPs, we only show the results for one wavelength (1.67$\,\mathrm{\mu m}$) and one position angle (along the inclination axis, corresponding to a pole on view.). The results are the same for the other spectral channels. We show the complete 1.67$\,\mathrm{\mu m}$ $V^2$ PIONIER dataset, but are only interested in fitting the short baseline data.
%
\section{Results and Discussion}
\label{sec:ResultsDiscussion}

\subsection{Radial flux distribution}
\begin{figure}
\centering
    \includegraphics[width=0.98\columnwidth]{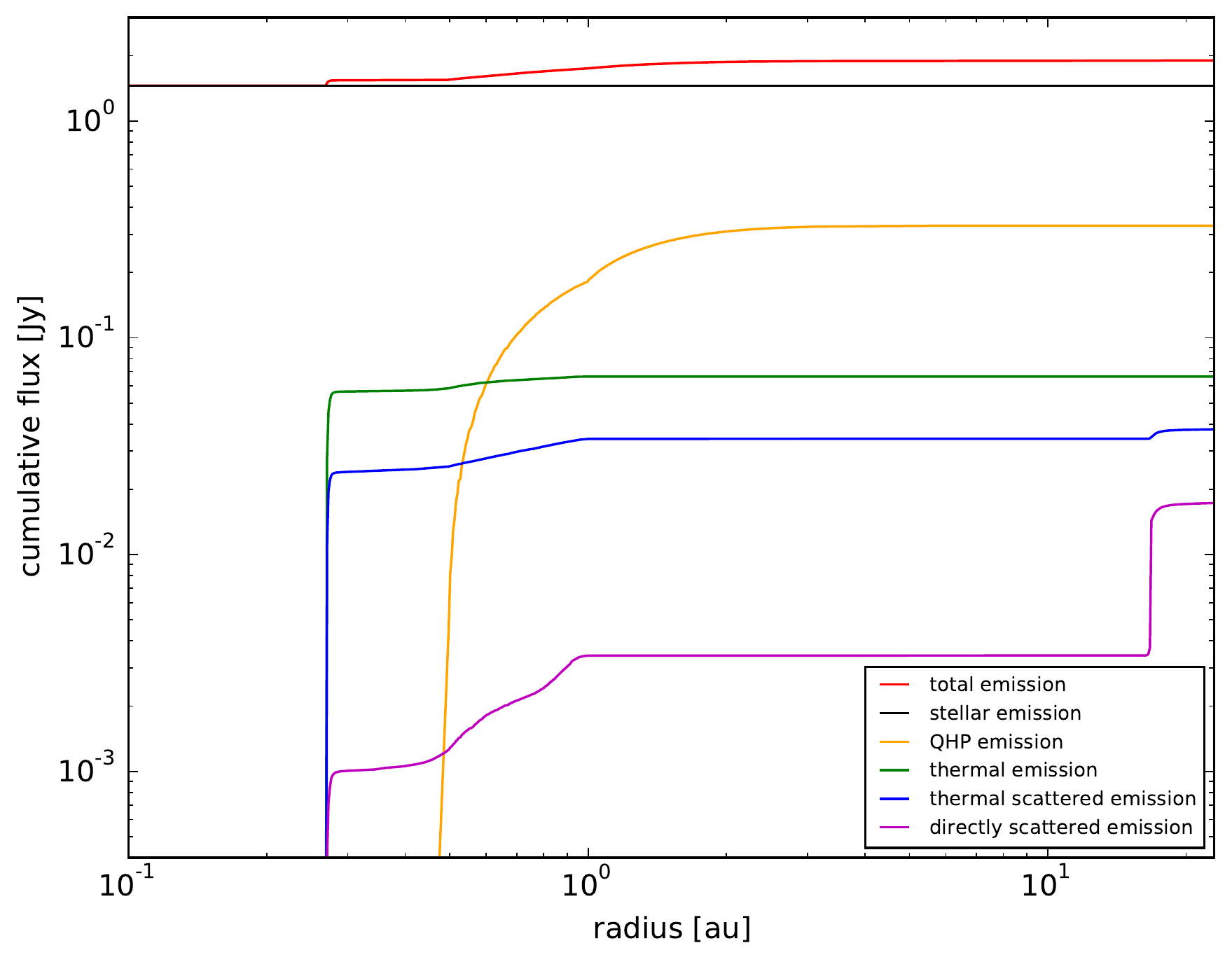}
     \caption{Cumulative H-band flux for model S18C100K, plotted against the distance from the star and split up into the different components: red:total emission; black:stellar emission; yellow:QHP emission; green:thermal emission; blue:thermal scattered emission; pink: directly scattered emission
}
      \label{fig:radial_flux}
\end{figure}

Fig.~\ref{fig:radial_flux} shows the radial distribution of the flux of model S18C100K at 1.67$\,\mathrm{\mu m}$ within the PIONIER field of view. The total flux is shown in red. The other lines show how different types of emission contribute to this total flux.\\
The stellar emission (black) contributes the majority of the total flux. The disk emission starts at the inner rim (0.27 au). It consists of three components: purely thermal emission (green), thermal emission from the disk that comes from photons that have been scattered at least once within the disk (blue) and directly scattered starlight (pink). In total, the purely thermal emission is about a factor three larger than the thermal scattered emission, which is again about a factor three larger than the emission from the directly scattered light. Thermal emission occurs at the inner rim and along the inner disk. The QHPs (yellow) start to contribute directly at 0.5\,au. After 2\,au, most of the UV photons have been absorbed and the QHPs at larger radii barely contribute additional flux. The outer disk is too cold to emit at 1.67$\,\mathrm{\mu m}$ (see Fig.~\ref{fig:denstemp}). Along the inner disk, the thermal scattered emission follows a similar profile as the thermal emission, but due to one (or more) scattering events, the outer disk also contributes. For the directly scattered light, the inner rim of the outer disk provides most of the flux, because it provides the largest scattering area.
\\
We compare this to the radial flux distribution P08 show for one of their models (their Fig.~1, upper panel). It shows the 2.2$\,\mathrm{\mu}$m emission of a Herbig star and its disk, but their star has a temperature of 10000$\,$K and  luminosity of 24$\,$L$_{\cdot}$. They use particles up to a size of 1$\,\mathrm{\mu}$m and with an albedo of 0.9 and assume an evaporation temperature of 2000$\,$ at 0.5$\,$au and do not include a gap or QHPs.
The shape of their disk emission curve is similar to the one shown in Fig.~\ref{fig:radial_flux}: a steep rise of the flux at the inner rim of the inner disk and additional emission along the inner disk. But there are three major differences between the radial flux distributions that can be explained by the different model setups. P08 differs inthe following ways:
\begin{itemize}
\item{The high disk temperature leads to a stronger contribution of the disk. The higher temperature and larger luminosity of the star do increase the stellar emission, but due to the shift of peak of the stellar emission, the amount of re-processed stellar emission growth stronger then the amount of direct stellar emission in the NIR. In addition, the emission area of the disk is larger since the disk starts at a larger radius. These effects lead to a stronger contribution of the disk compared to the stellar contribution.} 
\item{The overall hotter disk, in combination with the lower observation wavelength leads to more contribution along the inner disk than we find in our model.} 
\item{The higher albedo means that photons will scatter more often inside the disk, so the contribution of the thermal scattered emission is higher than the contribution from the thermal emission, while in our case it is the other way around.}
\end{itemize}


The QHP emission (yellow) follows a different radial profile. The emission rises slowly from 0.5$\,$ onwards and starts to dominate the disk emission from about 0.7$\,$au. It continues to rise outside of 1$\,$au in the area of the gap. In total, the QHPs contribute two times as much flux as the the combination of thermal and scattered light.\\
\begin{figure*}
\centering
    \includegraphics[width=0.98\columnwidth]{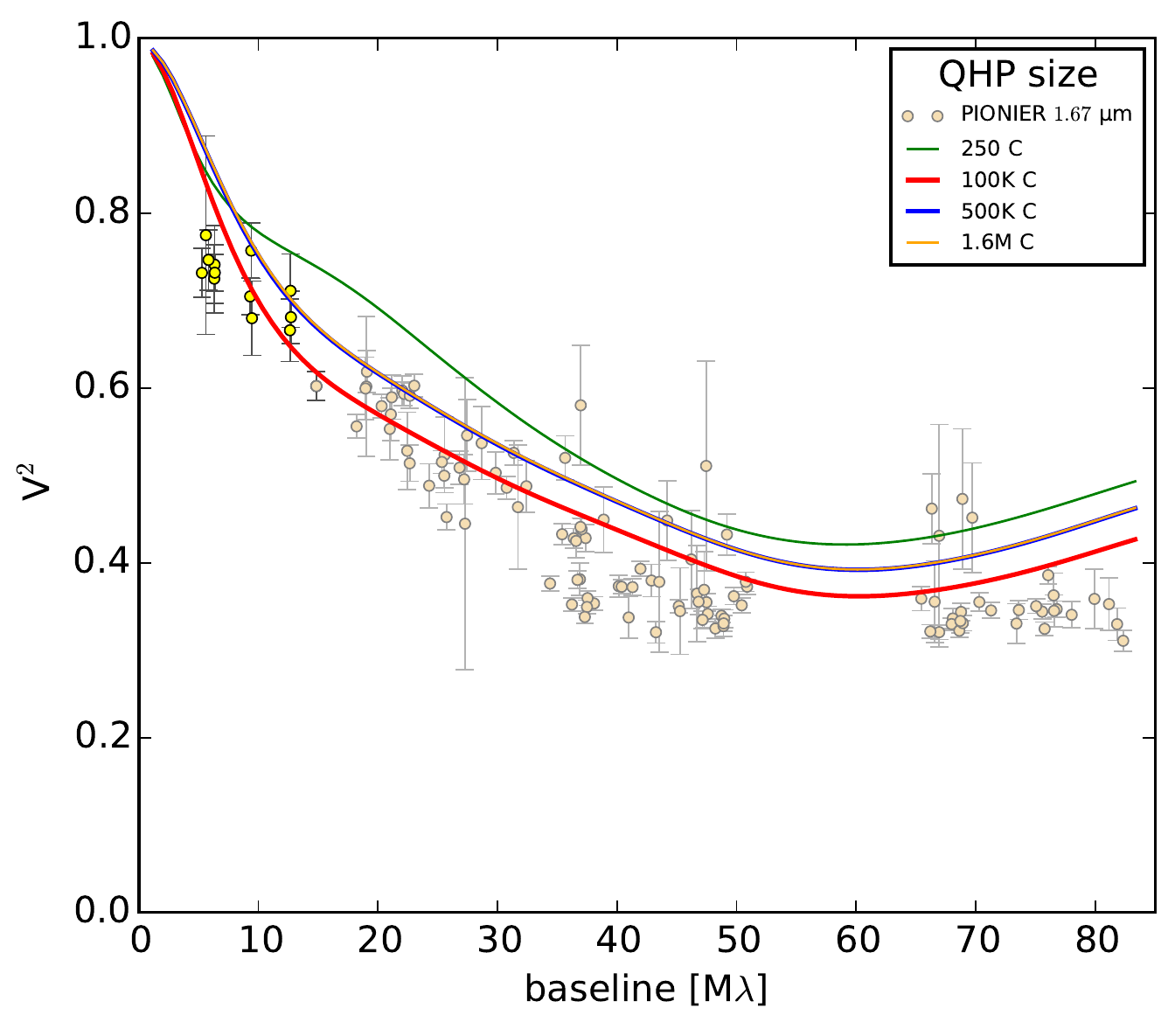}
    \includegraphics[width=0.98\columnwidth]{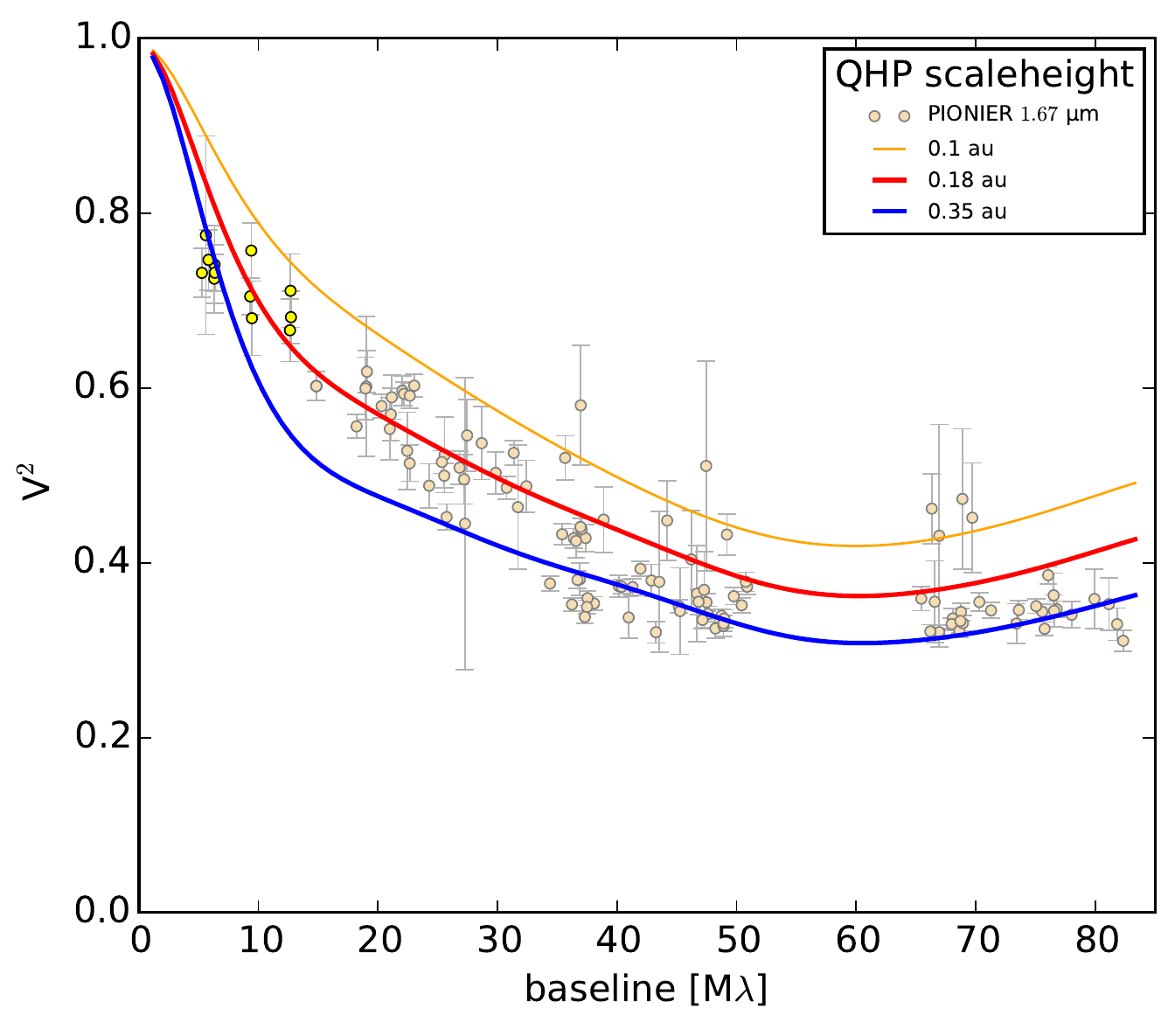}
    \includegraphics[width=0.98\columnwidth]{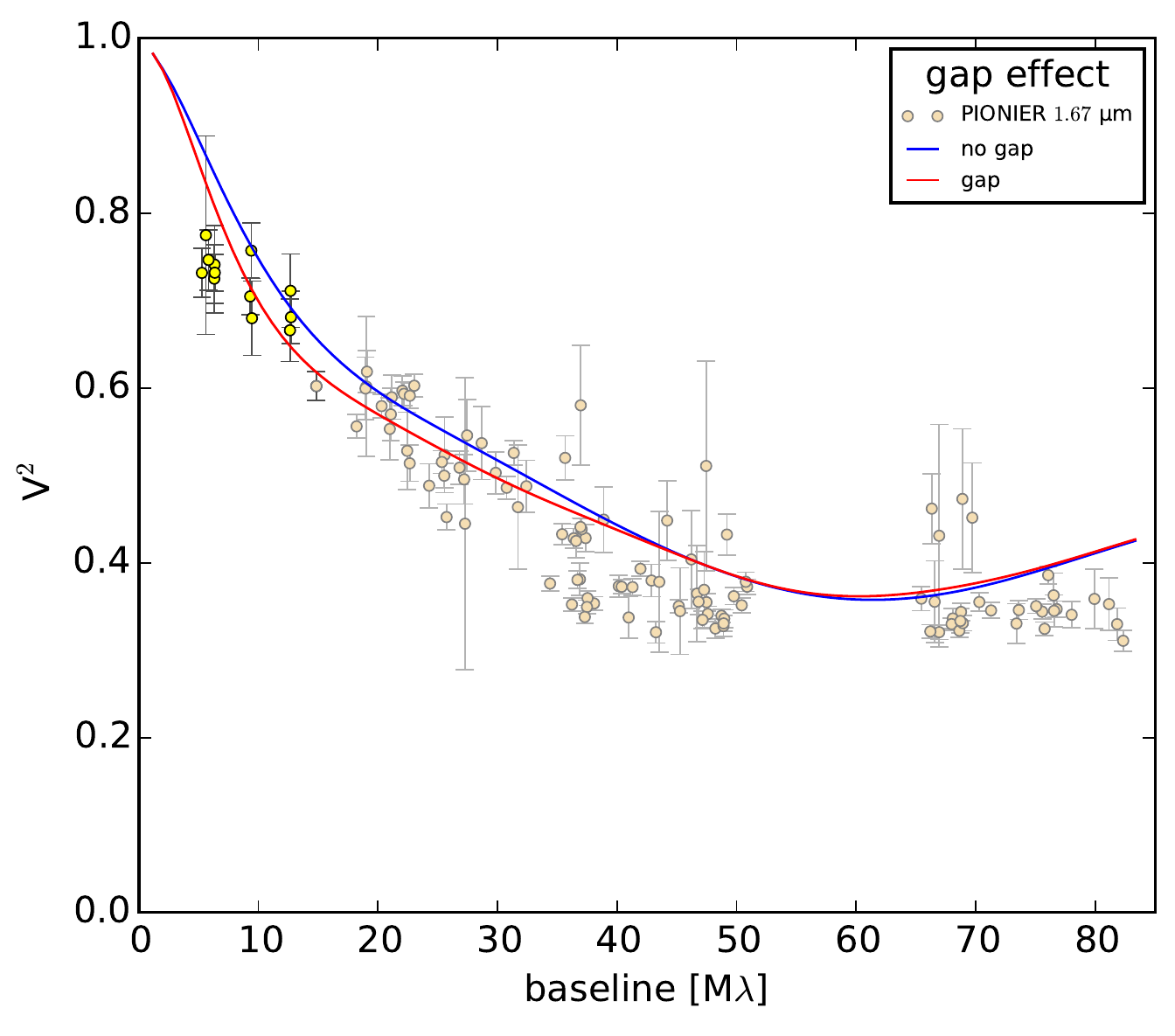}
    \includegraphics[width=0.98\columnwidth]{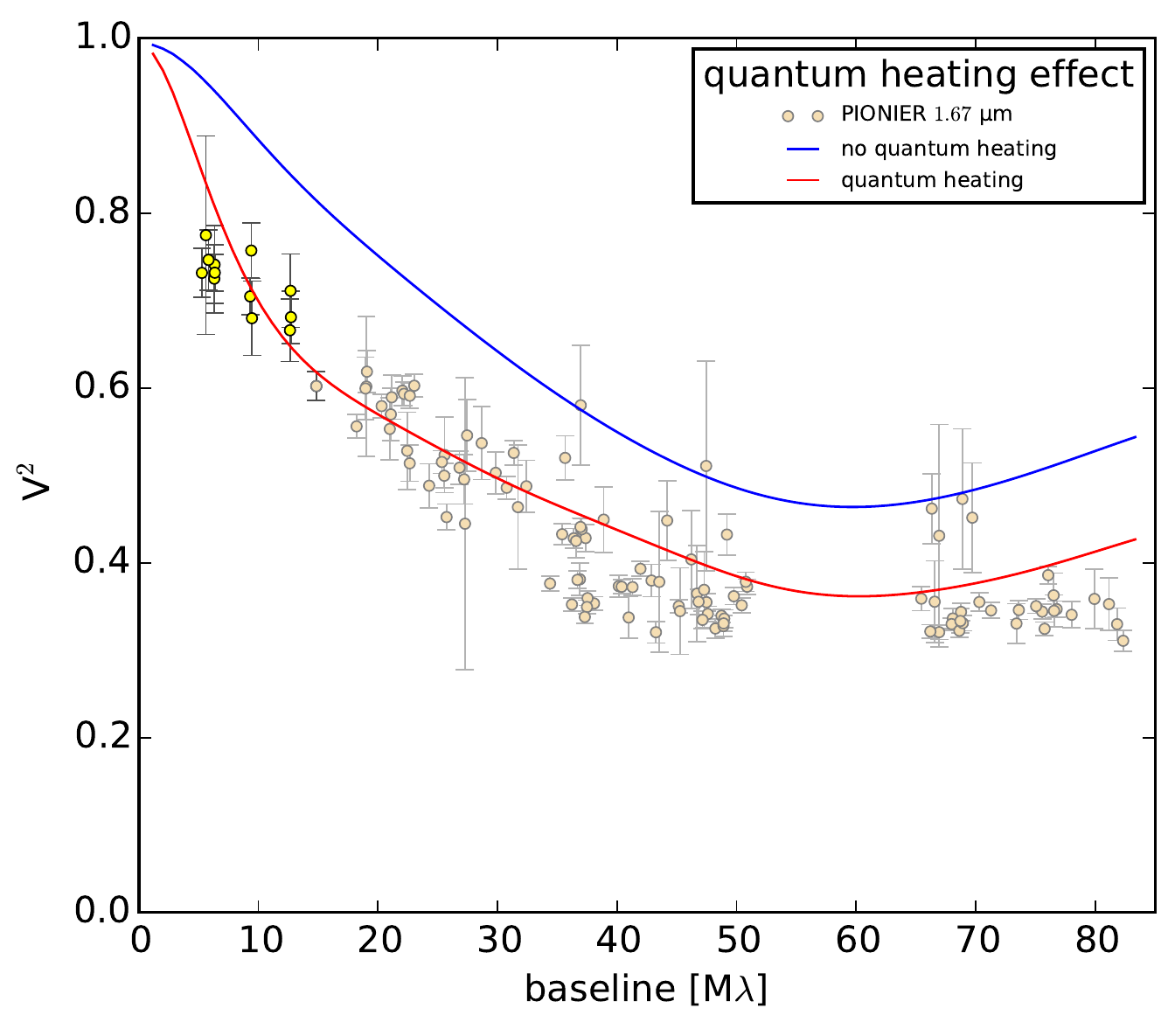}
     \caption{  
All panels: Squared, normalised visibilities $V^2$ against baseline in $\,$M$\mathrm{\lambda}$. The yellow circles show the 1.67$\,\mathrm{\mu m}$ data of HD$\,$100453 from the PIONIER Herbig Ae/Be survey \citep{Lazareff16, Kluska16}. For this analysis, only the short baselines are of interest (brightly colord symbols). The red line shows our best model, S18C100K. Top panels: We show the effect of different QHP sizes (green:S18C250, red:S18C100K, blue:S18C500K and yellow:S18C1.6M) and scale-heights (yellow:S10C100K, red:S18C100K, blue:S35C100K). Lower left panel: The QHPs lead to a (albeit smaller) drop even without the presence of a gap. Lower right panel: We show the effect of quantum heating. For the model corresponding to the blue line, S18C100Knqh, the grains behave like conventional dust. The drop at short baselines vanishes.
}
      \label{fig:Visibilities}
\end{figure*}

\subsection{QHP Size}
In this section, we investigate the influence of the QHP size on the Visibility curves and the PAH features.
\subsubsection{Visibilities}
In this section, we investigate the effects of the size of the QHPs. Looking first at the $V^2$ (Fig.~\ref{fig:Visibilities}, top left), the S18C250 model (green line) does not produce enough emission to create a deep enough drop. Model S18C100K with $10^{5}$ carbon atoms leads to the largest drop size (red line). While this model fits the 10$\,$M$\mathrm{\lambda}$ datapoints very well, it slightly overestimates the data at smaller baselines and slightly underestimates it at longer baselines. This could be an effect of our radial QHP distribution. Moving to even larger QHP sizes has the opposite effect and the drop depth becomes smaller again. It then stays the same for models S18C500K and S18C1.6M (blue and yellow line).\\
To demonstrate this we show a spectrum of the four models (Fig.~\ref{fig:Spectrum_size}). At 1.67$\,\mathrm{\mu m}$ (light yellow bar), model S18C100K contributes more flux then the models with QHPs of different sizes. Since we placed the QHPs at the correct spatial position (extended, but within 20 au), this is extended flux and creates the drop at short baselines. And since model S18C100K has the strongest flux it has also the largest drop.\\

\subsubsection{PAH features}
\begin{figure}
\centering
    \includegraphics[width=1.0\columnwidth]{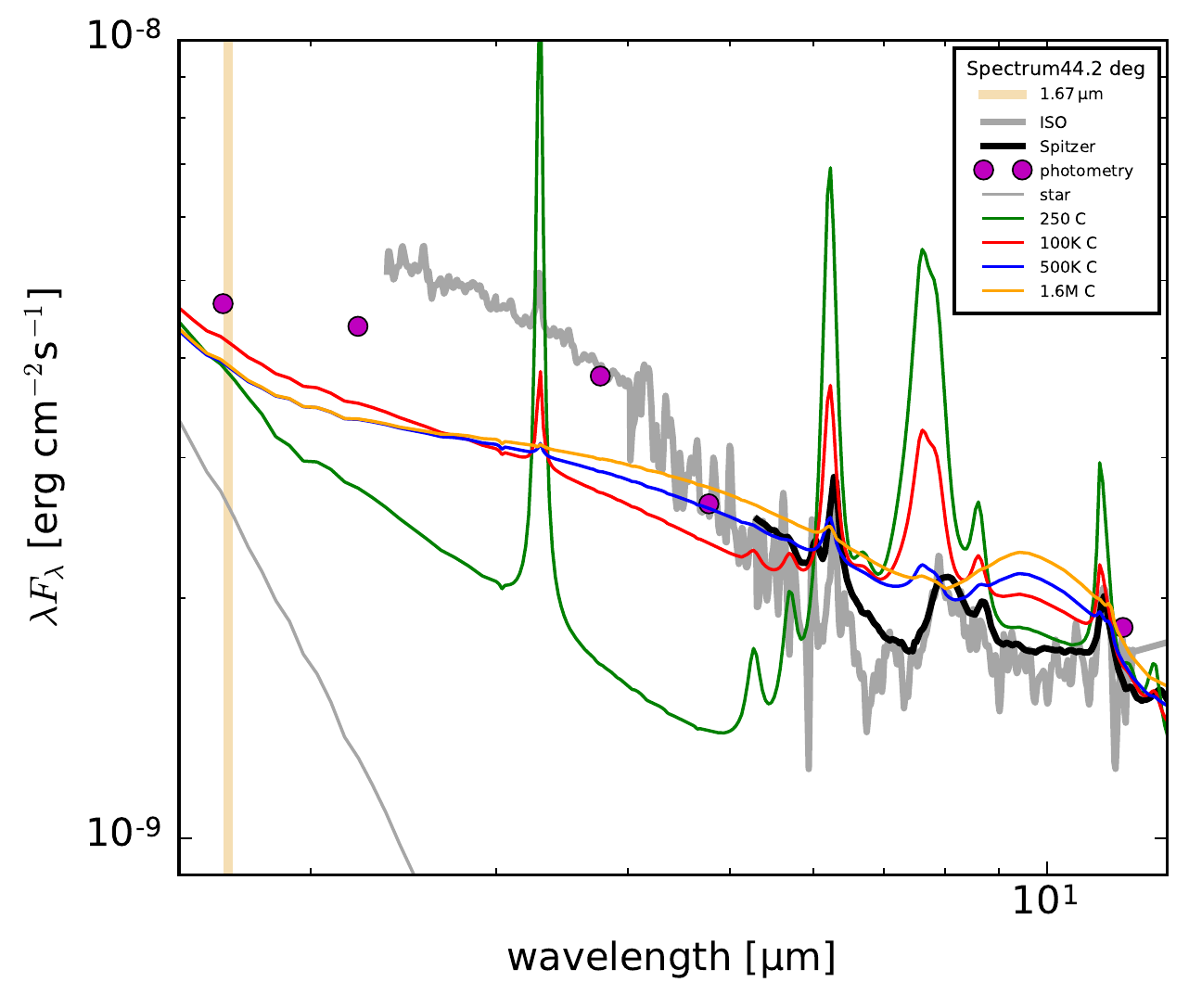}
     \caption{Influence of the QHP size on the strength of the PAH features. The ISO and Spitzer spectra are shown in gray and black. The photometry data is indicated by pink dots. A light yellow bar indicates 1.67$\,\mathrm{\mu m}$. The stellar spectrum is shown in light gray. The models have the same colors as in Fig.~\ref{fig:Visibilities}
      }
      \label{fig:Spectrum_size}
\end{figure}

Fig.~\ref{fig:Spectrum_size} also contains other observational data for HD$\,$100453: a Spitzer spectrum \citep[dark gray,][]{Acke10}, an ISO spectrum \citep[light gray,][]{Acke04} and photometry \citep[pink, K16, ][]{Malfait98}. Since there might be more QHPs in the outer region of the disk, the QHPs we added in the inner region should not overestimate the features too much. Model S18C250 (green) clearly overestimates each PAH feature. Model S18C100K (red) and model S18C500K (blue) are both borderline cases: Looking at the 6.2$\,\mathrm{\mu}$m feature (where the continuum emission of all models and the data are comparable), model S18C100K overestimates the flux in the feature by a factor 1.30, while S18C500K underestimates it by a factor 0.89. Looking at the the 3.3$\,\mathrm{\mu}$m feature, model S18C100K produces the correct feature size, while model S18C500K clearly underestimates it. Model  S18C1.6M underestimates all features.\\
To actually fit the PAH features it would be necessary to use different QHP size ranges and also consider different radial distributions (and QHPs outside of the PIONIER field of view), which is beyond the scope of this work. But we show that QHPs with the size of PAHs (up to a few hundred C atoms) clearly can not be the only source of the extended flux.\\
The ISO spectrum, the photometry data and the model are not in agreement with each other in the NIR (1-5$\,\mathrm{\mu}$m). Our visibilities have been observed at 1.67$\,\mathrm{\mu}$m, indicated by the light yellow column. The observed flux is slightly higher than the flux from model S18C100K. Looking at the visibility curve of this model in Fig.~\ref{fig:Visibilities} (e.g. lower right panel), the model also slightly overestimates the long baseline risibility data. This indicates that the missing flux is emitted on short spatial scales. To improve model S18C500K it would be necessary to modify the inner rim with a compact component, which is not part of this work.\\
Over the NIR range our models underestimate the flux by up to 30\%. This problem of the missing NIR flux is well known \citep[see for example][K16 and references therein]{Dullemond10}. While larger QHPs provide more NIR flux then PAHs with 250 C atoms, they still differ from the photometry data by a factor 1.25. Motivated by the missing compact flux at 1.67$\,\mathrm{\mu}$m we speculate that the same compact inner structure could provide the missing NIR flux. This approach is also taken in K16. But since the focus of this work is the extended flux and our observational wavelength is only marginally effected, we do not add a compact structure but use QHPs with 100K C atoms to explore other parameters that influence the extended flux.\\
\subsection{QHP Scaleheight}
\label{sub:QHP Scaleheigt}
In this section, we examine the influence of the QHP scale-height on the depth of the $V^2$ drop. Fig.~\ref{fig:Visibilities} (top right) shows the visibility curves for models with different QHP scahlheights: S10C100K (yellow), S18C100K (red) and S35C100K (blue). All models have a QHP size of $10^{5}$ carbon atoms. In model S35C100K, the QHPs are placed at a larger scale-height. That is why they intercept more light and contribute more flux at 1.67$\,\mathrm{\mu m}$, leading to a steeper drop. A scale-height of $0.1\,$au is not high enough to create a deep enough $V^2$ drop. The drop model S35C100K fits the datapoints at the smallest baselines well, but then the data is underestimated. When looking at the PAH features of this model (not shown in this paper), the model clearly overestimates the Spitzer data. We therefore use S18C100K as our best model.\\
A scale-height of $0.18\,$au correspond to a $z/r$ ratio of 0.18 at 1$\,$au and a $z/r$ ratio of about 0.01 at 17$\,$au. Especially the first value seems high for the dust scale-height of a disk. However, QHPs do track the gas distribution in the disk. In \citet{Woitke09} it can be seen that gas can extend up to a $z/r$ of 1 in this area of the disk. So we actually expect larger scale-heights for the gas in this area of the disk And while HD$\,$100453 is a transitional disk, the observations of the [OI] 63$\,\mathrm{\mu m}$ \citep{Meeus12,Fedele13} indicate that there is still some gas present.
\subsection{The Gap}
\label{sub:gap}
Due to the presence of the gap in HD$\,$100453, it is easier for QHPs to intercept UV photons, and the NIR photons emitted by the QHPs to escape the disk. In a disk without a gap, this effect might lead to less extended NIR flux and therefore to a smaller $V^2$ drop. To test the impact of the gap, we take model S18C100K and replace the gaped disk by a continuous one. As shown in Fig.~\ref{fig:Visibilities}, lower left panel, the drop of the continuous model S18C100Kng (blue) is only slightly smaller than the one of S18C100K (red). This is not surprising, since most of the flux comes from a scale-height that is above the self-consistently calculated scale-height of the dust disk.
\subsection{Quantum Heating}
Finally, we demonstrate that the extended NIR emission and the corresponding $V^2$ drop is indeed caused by QHPs, and not by scattered light from conventional grains of the same size. We therefore re-calculate model S18C100K, but treat the QHPs as if they are in thermal equilibrium. In Fig.~\ref{fig:Visibilities}, S18C100K corresponds to the red line, S18C100Knqh to the blue line. Without the quantum heating, the $V^2$ drop at short baselines vanishes.
%
\section{Conclusions and Future Research}
\label{sec:Conclusions}
We have shown basic models that demonstrate how the parameters of QHPs in protoplanetary disks can be constrained using NIR interferometry. A more detailed model of the complete PIONIER dataset of HD$\,$100453 that determines the flux contribution of each disk component in a more quantitative way is still necessary, especially a more detailed analysis of the position and shape of the inner rim, corresponding to the longer baselines. But these basic models already lead us to several conclusions: 
\begin{enumerate}
\item The PIONIER VLTI instrument has measured extended NIR emission from protoplanetary disks around Herbig stars, which leads to a $V^2$ drop at short baselines.
\item The flux in the PAH features from these disks is correlated with the depth of the $V^2$ drop, indicating that QHPs contribute significantly to the extended NIR emission.
\item This extended NIR emission can not be explained with thermal flux from grains in thermal equilibrium or scattered light. QHPs should therefore be considered when modelling NIR interferometry data of Herbig stars.
\item $10^{-\textbf{9}}\,$M$_{\odot}$ of QHPs containing $\sim10^{5}$ carbon atoms with a scale-height of 0.18$\,$au can produce the observed $V^2$ drop in HD$\,$100453 without deviating too much from to the PAH feature fluxes.
\item With detailed radiative transfer modelling it is possible to put constraints on the mass, size and position of the carbonaceous components in disks around Herbig stars using interferometric data.
\end{enumerate}
Aperture masking instruments like NACO at the VLT or NIRC2 at Keck observatory could help to gain more information about the distribution of the QHPs. New VLTI instruments like MATISSE and GRAVITY will allow an even more detailed analysis of protoplanetary disks around Herbig stars and could therefore contribute to the explanation of bulk carbon abundances of terrestrial planets around Herbig stars.\\

\paragraph{\textit{Acknowledgements}} The authors thank the anonymous referee for the comments and suggestions that helped to improve this paper. The authors are grateful to S. Khalafinejad and K. Maaskant for sharing their model parameters and discussing the modelling process. The authors thank K. Hakim, M. Kama and I. Kamp for helpful discussions. L.K. is supported by a grant from the Netherlands Research School for Astronomy (NOVA). L.K. acknowledges partial financial support from the Fizeau Exchange Visitor Program, funded by WP14 OPTICON/FP7 (grant number 312430). M.B. acknowledges financial support from "Programme National de Physique Stellaire" (PNPS) of CNRS/INSU, France. J.K. acknowledges support from a Marie Sklodowska-Curie CIG grant (Ref. 618910, PI: Stefan Kraus).
\bibliography{create_references}

\Online

\begin{appendix} 
\section{Density and Temperature structure}
\label{app:denstemp}
\begin{figure*}
\centering
    \includegraphics[width=2.0\columnwidth]{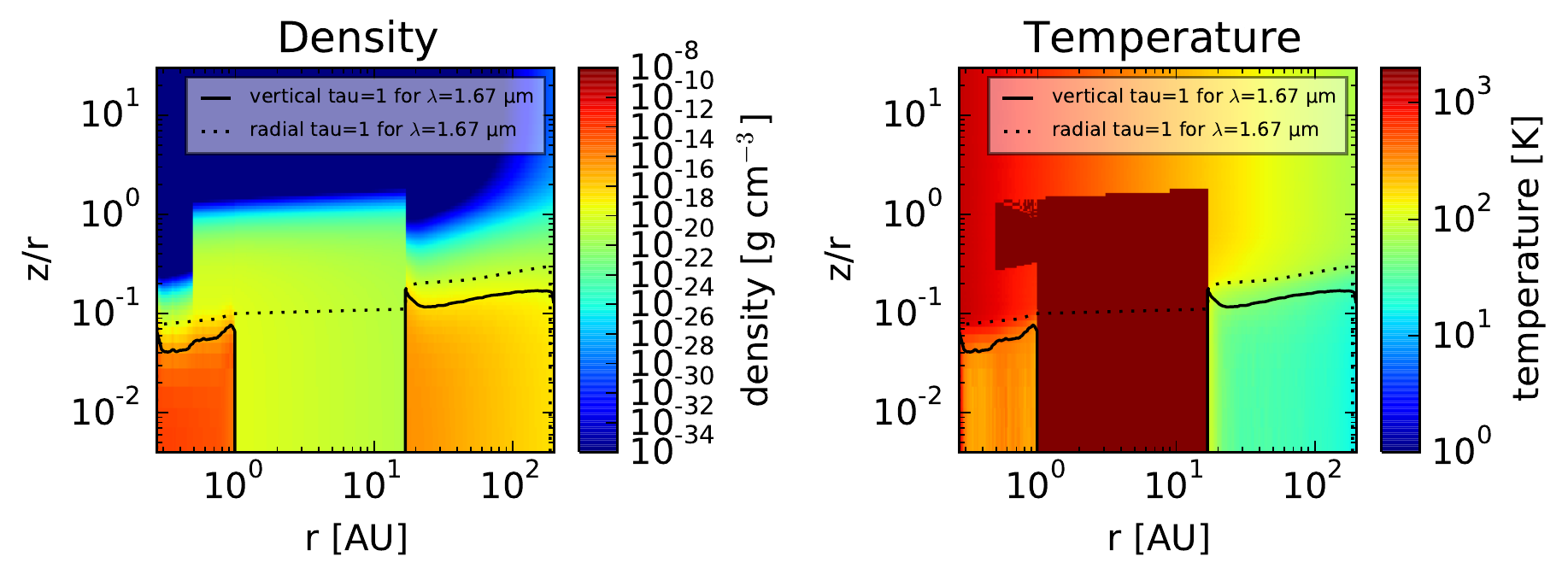}
    \includegraphics[width=2.0\columnwidth]{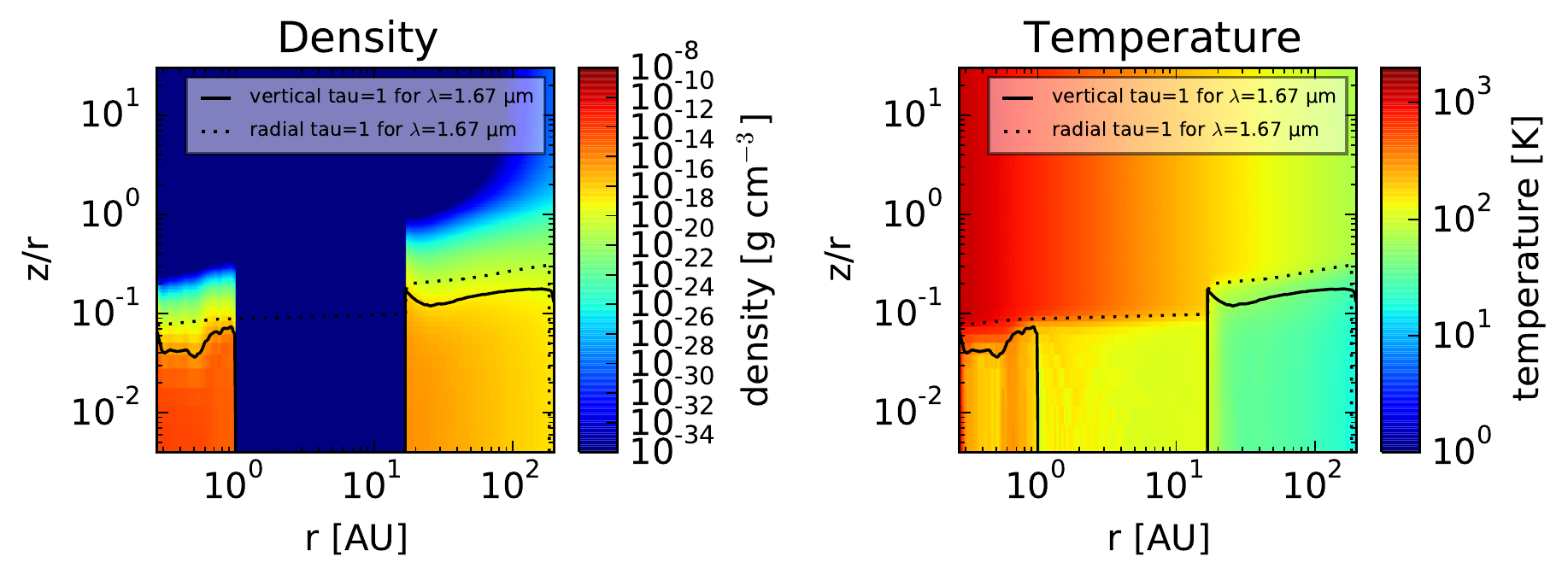}
     \caption{Final density and temperature structure for a model setup including (upper row, S18C100K) and without (lower row) QHPs. The x-axis shows the radius in au, the y-axis the height divided by the radius. The color of the left plots indicate the mass density, the colors of the right plot the temperature. The dark red area shows the regions with QHP emission. The optical depth at the observation wavelength is indicated by the radial (dotted line) and vertical (full line) $\tau=1$ surface.
      }
      \label{fig:denstemp}
\end{figure*}
Fig.~\ref{fig:denstemp} shows the final density and temperature structure of model S18C100K (upper row). The lower row shows the final structure of a model with the same setup as S18C100K, but without the QHPs. The QHPs have a mass density of about $10^{-20}\,\text{g}/\text{cm}^{3}$. For the dust in thermal equilibrium, we plot the equilibrium temperature. Only the inner disk is hot enough to thermally emit in the NIR. Areas that are dominated by QHP emission are colored in dark red.\\
The temperature distribution of the QHPs depends on their size and the strength of the local UV radiation field, which depends on the position of the QHPs. When a QHP is hit by a UV photon it is heated to very high temperatures, but cools down quickly by emitting NIR photons. This means that the QHPs follow a wide temperature distribution.\\
 In the shadow of the inner disk, most of them have a temperature of about  10\,K, but a small fraction reaches temperatures up to 2000\,K. In the well illuminated upper disk region, most of the QHPs have temepratures from a few hundred up to 2000\,K. QHPs of other sizes show a similar overall distribution, but with shifted temperature ranges. While QHPs with $1.6\times10^{6}$ carbon atoms reach 1700\,K, QHPs with only 250 carbon atoms reach up to 2400\,K.

\section{Flux table}

\begin{table*}
\caption{For each object shown in Fig. 2, we list the stellar flux $f_{\text{s}}$ , the extended flux $f_{\text{h}}$ and their $1-\sigma$ errors as calculated by \citet{Lazareff16} using a geometric model. We also give the size of the PIONIER field of view in au. The distances for each object have been taken from \citet{Acke04}.}
\label{tab:fluxes}
\centering
\begin{tabular}{lccccc}
\hline\hline
object & $f_{\text{h}}$ & $\sigma_{f_{\text{h}}}$ & $f_{\text{s}}$ & $\sigma_{f_{\text{s}}}$ & field of view [au] \\
\hline
HD 31648 &  0.01 &  0.008 &  0.37 &  0.050 &  26 \\
HD 34282 &  0.07 &  0.041 &  0.33 &  0.050 &  80 \\
HD 95881 &  0.04 &  0.005 &  0.27 &  0.022 &  24 \\
HD 97048 &  0.05 &  0.001 &  0.43 &  0.031 &  36 \\
HD 100453 & 0.11 &  0.005 &  0.58 &  0.002 &  23 \\
HD 100546 & 0.10 &  0.013 &  0.47 &  0.010 &  21 \\
HD 139614 & 0.06 &  0.009 &  0.54 &  0.018 &  28 \\
HD 141569 & 0.01 &  0.005 &  0.84 &  0.069 &  20 \\
HD 142527 & 0.03 &  0.008 &  0.42 &  0.011 &  29 \\
HD 144432 & 0.00 &  0.003 &  0.43 &  0.017 &  29 \\
HD 163296 & 0.01 &  0.005 &  0.25 &  0.009 &  24 \\
HD 169142 & 0.08 &  0.005 &  0.72 &  0.027 &  29 \\
MWC297 &    0.01 &  0.008 &  0.11 &  0.001 &  50 \\
HD 179218 & 0.03 &  0.021 &  0.58 &  0.003 &  48 \\
\hline
\end{tabular}
\end{table*}

\end{appendix}

\end{document}